\documentstyle[12pt]{article}
\begin{document}

\newcommand{\nonu}{\nonumber \\[2mm]}

\def\Tt{\tilde{T}}
\def\Wt{\tilde{W}}
\def\ct{\tilde{c}}
\def\bt{\tilde{b}}
\def\gat{\tilde{\gamma}}
\def\bet{\tilde{\beta}}
\def\p{\partial\phi_1}
\def\pp{\partial^2\phi_1}
\def\ppp{\partial^3\phi_1}
\def\pt{\partial\tilde\phi_1}
\def\ppt{\partial^2\tilde\phi_1}
\def\pppt{\partial^3\tilde\phi_1}
\def\jB{j_{{\rm BRST}}}
\def\la{\leftrightarrow}

\pagestyle{empty}
\rightline{UG-2/93}
%AS
\rightline{UCB-PTH-93/05}
%AS
\rightline{LBL-33737}
\rightline{March 1993}
\vspace{2truecm}
\centerline{\bf On the BRST Operator of $W$-Strings}
\vspace{2truecm}
\centerline{E.~Bergshoeff\footnote{
Institute for Theoretical Physics, Nijenborgh 4, 9747 AG Groningen,
The Netherlands.}, H.J.~Boonstra${}^1$, M.~de Roo${}^1$, S.~Panda${}^1$
and A.~Sevrin\footnote{
Theoretical Physics Group, Lawrence Berkeley Laboratory, 1 Cyclotron
Rd.~and Department of Physics, University of California at Berkeley,
Berkeley, CA 94720, USA.
}
}
\vspace{.5truecm}
\vspace{2truecm}
\centerline{ABSTRACT}
\vspace{.5truecm}
We discuss the conditions under which the BRST operator of a $W$-string can
be written as the sum of two operators that are separately nilpotent and
anticommute with each other. We illustrate our results with the example of the
non-critical $W_3$-string. Furthermore, we apply our results to make a
conjecture about a relationship between the spectrum of a non-critical
$W_n$-string and a $W_{n-1}$-string.

\vfill\eject
\pagestyle{plain}

\noindent{\bf 1. Introduction}

\vspace{.3cm}

In order to deal with a system of first-class constraints, it is convenient
to use the BRST formalism \cite{brst}. In particular, this approach
has turned out to be rather fruitful in the study of string theories
\cite{ka1,li1} and, more recently, in the study of critical
\cite{po1,po2,fr1} and
non-critical \cite{le1,oog,go1,bo1} $W$-string theories. In the quantum case,
the physical states are defined as the cohomology classes of a
nilpotent BRST operator $Q_B$. In the case of strings and $W$-strings,
the determination
of these cohomology classes is a rather complicated task and, sofar,
has only been solved completely for critical \cite{ka1} and
non-critical \cite{li1} strings.

Recently, it has been pointed out \cite{po1} that, in the case of a critical
$W_3$-string, the relevant BRST operator can be written as the sum
of two operators that are separately nilpotent and anticommute
with each other, i.e.
%AS
\begin{eqnarray}
\label{eq:split}
Q_B &=& Q_0+Q_1\nonumber\\
Q_0^2&=&Q_1^2=Q_0Q_1+Q_1Q_0=0
\end{eqnarray}
One thus ends up with a so-called {\sl double complex} which leads to a
double grading $(m,n)$ such that $Q_0$ has grading $(1,0)$ and $Q_1$
has grading $(0,1)$. One would expect that,
due to this additional structure, the determination of the
cohomology of $Q_B$ simplifies. For instance, each solution of the
$Q_0$ and $Q_1$ cohomology can now be used to construct a solution
of the full $Q_B$ cohomology. In view of these expected simplifications,
it is of interest to see whether the split-up of the BRST operator,
as given in (\ref{eq:split}), also occurs under more general
circumstances. It is the purpose of this paper to investigate
this issue. Our strategy will be to first discuss the generic
situation for classical systems. Next, we will consider the
quantisation. We will illustrate our results
with the example of the classical $w_3$-algebra and the recently
introduced modified $w_3$-algebra \cite{le1,be1}.
At the end of this letter we will
comment on the case of general $w_n$-algebras and make a conjecture
about the spectrum of the non-critical $W_n$-string.

\vspace{ .5truecm}
\noindent{\bf 2. The Classical case}

\vspace{.3cm}

In the classical BRST formalism, instead
of imposing constraints, one first extends
the standard phase space of coordinates and momenta with additional
anticommuting ghost coordinates. To be more precise, let us assume
that we have coordinates and momenta $z_A$ with canonical Poisson
brackets $\{z_A,z_B\}=\Omega_{AB}$. The matrix $\Omega$ defines a
symplectic structure on the phase space. Furthermore, let
$\Phi_\alpha(z_A)$ be a set of first-class constraints that form
the following Poisson-bracket algebra

\begin{equation}
\label{eq:coal}
\{\Phi_\alpha, \Phi_\beta\}=f_{\alpha\beta}{}^\gamma(z_A)\Phi_\gamma
\end{equation}

To each of the constraints we associate  a canonical pair of ghosts
$c^\alpha,b_\alpha$ with Poisson bracket $\{c^\alpha,b_\beta\}=-
\delta^\alpha{}_\beta$. In the BRST aproach one constructs,
corresponding to this system, a nilpotent BRST charge $Q_B$, i.e.~
$\{Q_B,Q_B\}=0$ (see e.g., \cite{he1}).
The generic structure of this BRST charge is given by

\begin{equation}
Q_B= c^\alpha\Phi_\alpha -{1\over 2} f_{\alpha\beta}{}^\gamma
c^\beta c^\alpha b_\gamma + \dots
\end{equation}
where the dots indicate terms of higher order in the ghosts.
If we assume that the structure constants satisfy the condition

\begin{equation}
\label{eq:jacid}
f_{[\alpha\beta}{}^\epsilon f_{\gamma]\epsilon}{}^\delta -
\{f_{[\alpha\beta}{}^\delta,\Phi_{\gamma]}\}=0
\end{equation}
then the BRST charge only contains terms linear and trilinear in the
ghosts. For simplictly, we will restrict our discussion to algebras
of this type. Let
$\Phi_\alpha(z_A) =\{T_a(z_A),W_i(z_A)\}$, then we also assume that in
terms of $T_a$ and $W_i$ the constraint algebra can be written as

\begin{eqnarray}
\label{eq:conal}
\{T_a,T_b\} &=& f_{ab}{}^c T_c \nonumber\\
\{T_a,W_i\} &=& f_{ai}{}^j(z_A) W_j \\
\{W_i,W_j\} &=& f_{ij}{}^k(z_A) W_k\,, \nonumber
\end{eqnarray}
where $f_{ab}{}^c$ is independent of $z_A$.
In particular, we note that both the $T_a$ and the $W_i$ form subalgebras
of the constraint algebra. Corresponding to the split of the generators,
it is natural to also split the ghosts as $c^\alpha=(c^a,\gamma^i)$ and
$b_\alpha=(b_a,\beta_i)$. The following theorem can now be proved.

\vspace{ .3truecm}

\noindent {\bf Theorem}\hskip .3truecm {\sl
The BRST charge $Q_B$ corresponding to the constraint algebra
(\ref{eq:conal}) can be written in the form

\begin{eqnarray}
\label{eq:clsplit}
Q_B&=&Q_0+Q_1\nonumber\\
\{Q_0,Q_0\}&=&\{Q_1,Q_1\}=\{Q_0,Q_1\}=0
\end{eqnarray}
with}
\begin{eqnarray}
\label{eq:clsplit2}
Q_0 &=& c^aT_a -{1\over 2} f_{ab}{}^c c^b c^a b_c - f_{ai}{}^j\gamma^ic^a
\beta_j\nonumber\\
Q_1 &=& \gamma^iW_i - {1\over 2}f_{ij}{}^k\gamma^j\gamma^i\beta_k
\end{eqnarray}

\vspace{ .5truecm}

The proof of this theorem is rather simple and relies on the fact that,
due to the particular form of the constraint algebra (\ref{eq:conal}),
the Jacobi identity
(\ref{eq:jacid}) decomposes into three independent sets of identities
\begin{eqnarray}
f_{[ab}{}^ef_{c]e}{}^d &=& 0\nonumber\\
f_{ab}{}^df_{id}{}^j - 2f_{ai}{}^kf_{bk}{}^j + 2\{f_{ai}{}^j,T_b\}&=&0\\
f_{[ij}{}^lf_{k]l}{}^m - \{f_{[ij}{}^m,W_{k]}\} &=&0\\
2f_{ai}{}^kf_{jk}{}^l + f_{ij}{}^kf_{ak}{}^l -2\{f_{ai}{}^l,W_j\}
-\{f_{ij}{}^l,T_a\}&=&0
\end{eqnarray}
which can separately be used to show that $\{Q_0,Q_0\}=0, \{Q_1,Q_1\}=0$ and
$\{Q_0,Q_1\}=0$, respectively.

\vspace {.3truecm}

\noindent{\bf 3. Examples}

\vspace{.5cm}

\noindent {\bf 1.} \hskip .3truecm
As a first example we discuss the classical $w_3$-algebra which
has a spin-2 generator $T(z)$ and a spin-3 generator $W(z)$. In
conformal OPE language the algebra takes the following form:
\begin{eqnarray}
\label{eq:w3}
T(z)T(w) &=& {2T(w)\over (z-w)^2}+{\partial T(w)\over z-w}+{\rm \ ...}\,,
\nonumber\\
T(z)W(w) &=& {3W(w)\over (z-w)^2}+{\partial W(w)\over z-w}+{\rm \ ...}\,, \\
W(z)W(w) &=& {2TT\over
 3(z-w)^2}+{\partial(TT)\over 3(z-w)}+{\rm \ ...}\,. \nonumber
\end{eqnarray}
A multi-scalar realisation of the algebra is given by \cite{ro1}
\begin {eqnarray}
\label{eq:ror}
T &=& -{1\over 2} (\p)^2 + T_X\,, \nonumber\\
W &=& {i\over 3\sqrt 6}\{ (\p)^3 + 6 (\p) T_X\}\,.
\end{eqnarray}
where $T_X$ is a multi-scalar realisation of the Virasoro algebra.
Note that
the $w_3$-algebra (\ref{eq:w3}) is not yet of the form (\ref{eq:conal})
and therefore we cannot apply our theorem. To bring the algebra in the
desired form we make the following redefinition of the generators:

\begin{eqnarray}
\Tt &=& T \nonumber\\
\Wt &=& W - {2i\over \sqrt 6}(\p)T = {4i\over 3\sqrt 6}(\p)^3
\end{eqnarray}
The algebra then becomes
\begin{eqnarray}
\label{eq:wt3}
\Tt(z)\Tt(w) &=& {2\Tt(w)\over (z-w)^2}+{\partial \Tt(w)\over z-w}+{\rm \
...}\,,
\nonumber\\
\Tt(z)\Wt(w) &=& {3\Wt(w)\over (z-w)^2}+{\partial \Wt(w)\over z-w}+{\rm \
...}\,, \\
\Wt(z)\Wt(w) &=&  \,{-2i\sqrt 6 \p
 \Wt \over (z-w)^2}+{-i\sqrt 6 \partial(\p \Wt)\over z-w}+{\rm \ ...}\,,
\nonumber
\end{eqnarray}
which is indeed of the form (\ref{eq:conal}). Applying our
theorem, we can now write the
BRST charge $Q_B=\oint {dz\over 2\pi i}j_B$
in the form (\ref{eq:clsplit}), (\ref{eq:clsplit2})
with the BRST currents $j_0$ and $j_1$
corresponding to $Q_0$ and $Q_1$, respectively, given by

\begin{eqnarray}
\label{eq:q1q2}
 j_0 &=& c\{\Tt + T_{(\gamma,\beta)} + {1\over 2}T_{(c,b)}\}\,,\nonumber \\
j_1 &=&\gamma\{\Wt- i\sqrt 6
\p\partial\gamma\beta\}\,.
\end{eqnarray}
Here $(c,b)$ and $(\gamma,\beta)$ are the usual spin-2 and spin-3 ghost
fields whose energy-momentum tensors are given by
\begin{eqnarray}
\label{TG2}
  T_{(c,b)} &=& -2b\partial c- \partial b c \,,\\
\label{TG3}
  T_{(\gamma,\beta)} &=& -3\beta\partial\gamma-2\partial\beta\gamma \,,
  \end{eqnarray}

\vspace{ .3truecm}

\noindent {\bf 2.} \hskip .3truecm As a second example we consider the
modified $w_3$-algebra of \cite{le1,be1}.
Let \{$T_M,W_M$\} and \{$T_L,W_L$\} be two commuting copies of
the classical $w_3$-algebra.
For the matter sector we again use the realisation
(\ref{eq:ror})\footnote{
Note that, instead of using realisation (\ref{eq:ror}) of the matter sector,
we could also use such a realisation for the Liouville sector. One could
then perform the same analysis as given below leading to the
same results with everywhere matter replaced by Liouville.
}.

The matter and Liouville parts can be combined
 into a single modified $w_3$-algebra
by defining the new generators
 $T=T_M +T_L$ and $W=W_M + i W_L$ \cite{le1,be1}.
This algebra then takes the form of a modified $w_3$-algebra  where, instead
of the third line of (\ref{eq:w3}), we have:
\begin{equation}
W(z)W(w) = {2(T_M-T_L)T\over
  3(z-w)^2}+{\partial\{(T_M-T_L)T\}\over 3(z-w)}+{\rm \ ...}\,.
\end{equation}
Note that $T_M-T_L$ appears as a structure function, i.e.~the algebra is of the
so-called soft type.

Again, a redefinition of the generators can be made such that the algebra
is of the form (\ref{eq:conal}). In the present case, this redefinition
is given by

\begin{eqnarray}
\Tt &=& T\nonumber\\
\Wt &=& W - {2i\over \sqrt 6}(\p)T= {4i\over 3\sqrt 6}(\p)^3
-{2i\over \sqrt 6}\p T_L+iW_L
\end{eqnarray}
In terms of the new generators the algebra is
given by the same formula (\ref{eq:wt3}) which we already encountered
in the case of the unmodified $w_3$-algebra. The same applies to the BRST
charge. It is again given by eq.~(\ref{eq:q1q2}).

As a byproduct of our manipulations, we see that, by allowing
realization-dependent redefinitions of the generators, the classical
$w_3$-algebra and the modified $w_3$-algebra can be brought into exactly the
same form as
given in (\ref{eq:wt3}). Of course, the {\sl realization} of the generators
is different in both cases.
%AS
Obviously, the previous example is contained in this example as can be seen by
putting $T_L=W_L=0$.

\vspace {.3truecm}

\noindent{\bf 4. Quantisation}

\vspace{.5cm}

To describe the quantisation, it is convenient to use the Batalin-Vilkovisky
(BV) formalism \cite{bv}. This formalism enables one, under the condition that
the theory has no anomalies, to derive from a given classical
BRST charge a nilpotent BRST operator by means of an iterative procedure.
This BV formalism has for instance been applied in \cite{be1}
to derive the quantum
BRST operator corresponding to the modified $w_3$-algebra. A priori, it is not
guaranteed by this procedure that a classical BRST charge with the
property (\ref{eq:clsplit}) carries over into a quantum BRST operator with
the property (\ref{eq:split}). However, in all cases we have investigated,
the property (\ref{eq:split}) turns out to hold.

The expression for the quantum BRST operator corresponding to the
redefined $w_3$-algebra can be
found in \cite{po1} so we will not repeat it here.
In \cite{po1} it was noted that the quantum BRST operator corresponding
to the redefined algebra (\ref{eq:wt3}) can be related to the quantum BRST
operator corresponding to the unmodified $w_3$-algebra by means of a
field redefinition of the ghosts and the matter fields. In case
of the redefined modified $w_3$-algebra, instead of deriving the quantum
BRST operator through the BV formalism, we found it much simpler to
use the quantum BRST operator corresponding to the modified
$w_3$-algebra given in \cite{le1} and to apply a similar redefinition on it.
Of course, the result is the same as what would follow from the
BV formalism.

We have found that the quantum counterparts of the classical BRST currents
(\ref{eq:q1q2}) are given by
\begin{eqnarray}
\label{J1}
 j_0 &=& c\,\{ T_{\phi_1} + T_X + T_L
     + T_{(\gamma,\beta)} +{1\over 2} T_{(c,b)} \} \,,
  \\
 j_1 &=&\gamma\,\big[\,
  {i\over 3\sqrt{6}}\,\{
   4(\p)^3 - 12q_1\p\pp + (-15+4q_1^2)\ppp \}
  \nonumber\\
  &&\qquad + i\,\{W_L- {2\over \sqrt{6}}\p T_L +{1\over \sqrt{6}}q_1
  \partial T_L\}
   \nonumber\\
  \label{J2}
  &&\quad -i\sqrt 6\{ \p\partial\gamma\beta
+{1\over 3}q_1\partial\beta\partial\gamma\}\big] \,.
\end{eqnarray}
The quantum energy momentum tensor for the $\phi_1$ scalar is given by
\begin{equation}
T_{\phi_1} = -{1\over 2}(\p)^2 + q_1\partial^2 \phi_1
\end{equation}
This scalar contributes
$c_{\phi_1}= 1+12q_1^2$ to the
total central charge, while the additional matter scalars, with energy-momentum
tensor $T_X$, contribute $c_X=1+4q_1^2$ \cite{ro1}.
The total central charge contribution of the matter and Liouville
sectors is thus
given by $c_M=c_{\phi_1}+c_X=2+16q_1^2$ and $c_L$, respectively.
Furthermore,
the energy-momentum tensors $T_{(c,b)}$
and $T_{(\gamma,\beta)}$ of the spin-2 and spin-3 ghosts
satisfy a Virasoro algebra  with
central charge $-26$ and $-74$, respectively.
Finally, the generators $(T_L,W_L)$ satisfy a
quantum $W_3$-algebra with central charge $c_L$.
The nilpotency of the BRST operator requires that $c_M+c_L=100$.

\vspace {.3truecm}
\vfill\eject
\noindent{\bf 5. Comments}

\vspace{.5cm}

It has been suspected for a long time that various non-critical
$W_n$ strings are somehow related.  Evidence supporting this
can be found in the appearence
of $W_n$ constraints in the double  scaling limit of multimatrix models and
in the results for the spectrum of pure $W_3$-gravity \cite{da1,ram}.
We believe
that the present work strongly suggests a larger scheme which would enable one
to establish relations across models.

In order to formulate our conjecture, we make two observations. In
\cite{da1,po3},
it was shown that a set of $W_{n-1}$ currents and one free scalar field
with a fixed background charge allows one to construct the $W_n$ currents. A
second point is that the work in \cite{le1,be1} very probably generalizes to
arbitrary $W_n$ models in the following way. Consider two mutually
commuting copies of
the $W_n$ algebra, generated by $T_M^{(j)}$ and $T_L^{(j)}$ resp., where $2\leq
j\leq n$. The total currents $T^{(j)}$ are
$T^{(j)}\equiv T_M^{(j)}+ i^{j-2}T_L^{(j)}$\footnote{This observation was made
in collaboration with X.~Shen.}.
We next write the Liouville currents in terms of a scalar field $\phi$ and
$W_{n-1}$ currents\footnote{
This is an alternative to what we did in example 2 of section 3 where
we specified a realization of the matter sector. See also the footnote
in that section.}. We conjecture that the BRST operator for this system can
be written in the form (\ref{eq:split}) where $Q_0$ is the BRST operator for
a $W_{n-1}$ system.
In $Q_0$, the ghosts for the
spin $n$ symmetry together with the scalar $\phi$ have become ``matter'' for
the $W_{n-1}$ system. This conjecture can presumably be proven by combining the
two observations above.

The following counting argument supports our conjecture.
Consider $W_{n}$ gravity coupled to $(p,q)$ $W_n$ minimal
matter, {\it i.e.} non-critical $W_n$ strings.
The central charge $c_M$ of the $(p,q)$ minimal model is given by
\begin{equation}
c_M=(n-1)(1-n(n+1)Q_M^2),
\end{equation}
where
\begin{equation}
Q_M=\sqrt{\frac p q}-\sqrt{\frac q p}\,.
\end{equation}
The $W_n$ gravity sector is an $Sl(n,{\bf R})$ Toda system with central charge
$c_T$
\begin{equation}
c_T=(n-1)(1+n(n+1)Q_T^2),
\end{equation}
where
\begin{equation}
Q_T=\sqrt{\frac p q}+\sqrt{\frac q p}.
\end{equation}
The two central charges add up to
\begin{equation}
c_M+c_T=2(n-1)(2n^2+2n+1)
\end{equation}
which gets precisely cancelled by the $W_n$ ghost contributions to the total
$c$. Following \cite{da1,po3}, the $W_n$ Toda currents can now be rewritten in
terms of $W_{n-1}$ currents and one scalar field $\phi$, with central charge
$c_{\phi}=1+3 n (n-1)Q_T^2$. If our conjecture is true, the $W_n$ minimal
matter, together with $\phi$ and the ghost and the anti-ghost of the spin $n$
symmetry (which has a central charge $c_{(c^{(n)},b^{(n)})}=-2(6n^2-6n+1)$),
combine to a matter sector with $W_{n-1}$ symmetry. This matter has central
charge:
\begin{equation}
\tilde{c}_M=c_M+c_{\phi}+c_{(c^{(n)},b^{(n)})}
=(n-2)(1-n(n-1)Q_M^2)
\end{equation}
which is precisely what we expect for the $(p,q)$  $W_{n-1}$ minimal model.
Our conjecture suggests novel realizations of $(p,q)$ $W_n$ minimal models in
terms of $n+k-1$ scalar fields and $k$ $b-c$ systems with $1\le k\le n-2$.

If we restrict ourselves to unitary minimal models, {\it i.e.}
choosing $q+1=p$, we get that the following relation across lines should
exist:

\begin{equation}
\begin{array}{lclclc}
\mbox{Pure }W_2     &   &                   &   &               & \\
\mbox{gravity}      &   &                   &   &               & \\
                    &   &                   &   &               & \\
\mbox{Ising }(c=1/2) &\la&\mbox{Pure }W_3    &   &               & \\
+ W_2 \mbox{ gravity}  &   &\mbox{gravity}     &   &               & \\
                    &   &                   &   &               & \\
\mbox{tricritical}  &\la&\mbox{3-State}      &\la&\mbox{Pure }W_4& \\
\mbox{Ising}\ (c=7/10)&   &\mbox{Potts}\ (c=4/5)&   &\mbox{gravity} & \\
+ W_2\mbox{ gravity}  &   &+W_3\mbox{ gravity} &   &               & \\
                    &   &                   &   &               & \\
\vdots              &   &\vdots             &   &\vdots         & \ddots
\end{array}
\end{equation}
Our conjecture amounts to the claim that the spectrum of every model in the
table contains as a subsector the spectrum of the model to its left. In other
words, a $(p,q)$ non-critical $W_n$ string contains in its spectrum the
spectrum of $(p,q)$ non-critical $W_{n-k}$ strings where $1\leq k\leq n-2$.

Repeated applications of our conjecture
would result in the factorization of Q into a Virasoro BRST charge and
$n-2$ other, mutually anticommuting, nilpotent charges. This is
supported by the fact that {\sl any} classical $w_n$-algebra can be
realised in terms of an arbitrary number of scalars $X^\mu$ and $n-2$
special scalars $\phi_i\ (i=1,2,\dots, n-2)$ as follows:
\begin{eqnarray}
T&=&T_X + \sum_{i=1}^{n-2} T_{\phi_i}\nonumber\\
&\vdots&\nonumber\\
W^{(n-3)}&\sim& \sum_h (\partial\phi_{(n-3)})^{n-1-h}
                        (\partial\phi_{(n-2)})^{h} \nonumber\\
W^{(n-2)}&\sim& (\partial\phi_{(n-2)})^n \,.
\end{eqnarray}
The generic structure of a $w_n$-algebra in this basis is of the form

\begin{eqnarray}
\label{eq:wn}
{[}T,T{]} &\rightarrow&  T \nonumber\\
&\vdots& \nonumber\\
{[}W^{(n-3)},W^{(n-3)}{]}&\rightarrow& W^{(n-3)},W^{(n-2)} \nonumber\\
{[}W^{(n-3)},W^{(n-2)}{]}&\rightarrow& W^{(n-3)},W^{(n-2)} \nonumber\\
{[}W^{(n-2)},W^{(n-2)}{]}&\rightarrow& W^{(n-2)} \,.
\end{eqnarray}
This generalizes the structure given in (\ref{eq:conal}). It
would be interesting to see whether an algebra of the type
(\ref{eq:wn}) allows a multi-split of the corresponding BRST charge
of the form

\begin{eqnarray}
Q_B &=& \sum_{i=0}^{n-2} Q_i\nonumber\\
\{Q_i,Q_j\}&=&0
\end{eqnarray}
One could even consider taking the limit $n\rightarrow
\infty$ and construct a $W_\infty$-string theory.
More details on the generic case of classical $w_n$ algebras and their
quantisation will be given elsewhere \cite{be2}.

\vspace{ 1.3truecm}
\vfill\eject
\centerline{\bf Acknowledgements}

\vspace{.5truecm}

The work of E.B.~has been made possible by a fellowship of the
Royal Netherlands Academy of Arts and Sciences (KNAW).
The work of H.J.B.~and S.P.~was
performed as part of the research program of the
``Stichting voor Fundamenteel Onderzoek der Materie'' (FOM).
The work of
A.S. was supported in part by the Director Office of Energy Research
and Nuclear Physics, Division of High Energy Physics of the U.S.
Department of Energy under Contract DE-AC03-765F00098, and in part
by the National Science Foundation under Grant PHY90-21139.


\begin{thebibliography}{99}

\bibitem{brst}{ C.~Becchi, A.~Rouet and R.~Stora, Phys.~Lett.~B52
(1974) 344; I.V.~Tyutin, Lebedev preprint FIAN No.~39 (1975), unpublished.
}

\bibitem{ka1}{M.~Kato and K.~Ogawa, Nucl.~Phys.~B212 (1983) 443.
}

\bibitem{li1}{B.~Lian and G.~Zuckerman, Phys.~Lett.~254B (1991) 417;
Phys.~Lett.~266B (1991) 21; Comm.~Math.~Phys.~145 (1992) 561.
}

\bibitem{po1}{H.~Lu, C.N.~Pope, S.~Schrans and X.J.~Wang, ``On the
Spectrum and Scattering of $W_3$-Strings'', preprint CTP TAMU-4/93,
KUL-TF-93/2 (January 1993).
}

\bibitem{po2}{C.N.~Pope, L.J.~Romans and K.S.~Stelle, Phys.~Lett.~268B (1991)
167; C.N.~Pope, E.~Sezgin, K.~Stelle and X.J.~Wang, ``Discrete
States in the $W_3$ String'', preprint CTP TAMU-64/92; C.N.~Pope, ``Physical
States in the $W_3$ String'', preprint CTP TAMU-71/92; H.~Lu, B.E.W.~Nilsson,
C.N.~Pope, K.S.~Stelle and P.C.~West, ``The Low-level Spectrum of the $W_3$
String'', preprint CTP TAMU-70/92; H.~Lu, C.N.~Pope, S.~Schrans and X.J.~Wang,
``The Interacting $W_3$ String'', preprint CTP TAMU-86/92.
}

\bibitem{fr1}{M.~Freeman and P.~West, ``$W_3$ String Scattering'',
preprint KCL-TH-92-4; P.~West, ``On the Spectrum, no ghost theorem and
modular invariance of $W_3$ Strings'', preprint KCL-TH-92-7; M.~Freeman
and P.~West, ``The Covariant Scattering and Cohomology of $W_3$
Strings'', preprint KCL-TH-93-2
}

\bibitem{le1}{M.~Berschadsky, W.~Lerche, D.~Nemeschansky and N.P.~Warner,
Phys.~Lett.~B292 (1992) 35; ``Extended $N=2$ Superconformal Structure of
Gravity and $W$-Gravity Coupled To Matter'', preprint CERN-TH.6694/92;
W.~Lerche,``Chiral Rings in Topological $(W-)$ Gravity'', preprint
CERN-TH.6812/93.
}

\bibitem{oog} H.~Ooguri, K.~Schoutens, A.~Sevrin and P.~van Nieuwenhuizen,
Comm.~Math.~Phys.~145 (1992) 515; K.~Schoutens, A.~Sevrin and P.~van
Nieuwenhuizen, Nucl.~Phys.~B371 (1992) 315.


\bibitem{go1} J.~de Boer and J.~Goeree, ``Covariant $W$-gravity and
its Moduli Space from Gauge Theory'', preprint THU-92/14; J.~Goeree,
``Higher-spin Extensions of Two-dimensional Gravity'', Ph.D.~Thesis,
Utrecht University.

\bibitem{bo1}{P.~Bouwknegt, J.~McCarthy and K.~Pilch, ``Semi-infinite
Cohomology of $W$-Algebras'', preprint USC-93/11.
}

\bibitem{he1}{M.~Henneaux, Physics Reports 129 (1985) 1; J.~Govaerts,
``Hamiltonian Quantisation and Constrained Dynamics'', Vol.~4, series B:
Theoretical Particle Physics, Leuven University Press.
}

\bibitem{be1}{E.~Bergshoeff, A.~Sevrin and X.~Shen, Phys.~Lett.~B296 (1992)
95.
}

\bibitem{ro1}{L.~Romans, Nucl.~Phys.~B352 (1991) 829.
}

\bibitem{bv}{J.A. Batalin and G.A. Vilkovisky, Phys.~Rev.~D28 (1983) 2567;
D30 (1984) 508; Nucl.~Phys.~B234 (1984) 106.
}


\bibitem{da1} S.~Das, A.~Dhar and S.K.~Rama, Int.~J.~Mod.~Phys.~A7 (1992)
2295.

\bibitem{ram} S. K. Rama, Mod. Phys. Lett.~6 (1991) 3531.

\bibitem{po3} H.~Lu, C.N.~Pope, S.~Schrans and K.W.~Xu, Nucl.~Phys.~B385
(1992) 99; H.~Lu and  C.N.~Pope, Phys.~Lett.~286B (1992) 63.

\bibitem{be2} E.~Bergshoeff, H.J.~Boonstra, M.~de Roo, S.~Panda, A~Sevrin
and X.~Shen, in preparation.

\end{thebibliography}
\end{document}